# Reduction of Hysteresis for Carbon Nanotube Mobility Measurements Using Pulsed Characterization

David Estrada,[1] Sumit Dutta,[1] Albert Liao,[1] and Eric Pop[1,2,*]

[1]*Dept. of Electrical and Computer Engineering, Micro and Nanotechnology Laboratory, University of Illinois, Urbana-Champaign, IL 61801, USA*

[2]*Beckman Institute, University of Illinois, Urbana-Champaign, IL 61801, USA*

**Abstract:** We describe a pulsed measurement technique to suppress hysteresis for carbon nanotube (CNT) device measurements in air, vacuum, and over a wide temperature range ($80 - 453$ K). Varying the gate pulse width and duty cycle probes the relaxation times associated with charge trapping near the CNT, found to be up to the $0.1 - 10$ s range. Longer off times between voltage pulses enable consistent, hysteresis-free measurements of CNT mobility. A tunneling front model for charge trapping and relaxation is also described, suggesting trap depths up to $4 - 8$ nm for CNTs on $SiO_2$. Pulsed measurements will also be applicable to other nanoscale devices such as graphene, nanowires, and molecular electronics, and could enable probing trap relaxation times in a variety of material system interfaces.

**\*Contact:** epop@illinois.edu

## 1. Introduction

Carbon nanotube field effect transistors (CNT FETs) are candidates for future nanoelectronics due to their ability to carry large current density and their high mobility, greater than $10^9$ A/cm$^2$ and $10^4$ cm$^2$/V·s respectively [1, 2]. In many studies, CNT FETs are grown or dispersed onto an insulator and back-gated by a silicon substrate. Hysteretic behavior in the drain current ($I_D$) with gate-to-source voltage ($V_{GS}$) transfer characteristics is often observed, and varies depending on sweep direction, sweep rate, and environmental conditions. This is typically attributed to charge trapping by surrounding water molecules or charge injection into the dielectric substrate [3-10]. Sweeping $V_{GS} > 0$ typically shifts the threshold voltage ($V_T$) up because of charge screening from injected electrons into trap sites. Similarly, sweeping $V_{GS} < 0$ induces hole injection into the CNT surrounding, and the threshold voltage is shifted down [11]. This leads to the observed "open eye" characteristics when continuous (DC) $I_D$-$V_{GS}$ measurements are made (see, e.g. Refs. 3-7 or figure 5), which cause uncertainty in measured threshold voltage, conductance, and mobility. In a DC sweep the charges remain trapped until the gate polarity is switched [12]. Although this hysteretic behavior can be exploited to create nonvolatile memory devices [11, 13, 14], it is often unclear which electrical characteristics should be used to extract carrier mobility and threshold voltage for transistor applications. This has lead to large discrepancies (> 10×) in reported mobility values as both the reverse [1] and forward [2] $I_D$–$V_{GS}$ sweeps have been used to extract mobility, and in some studies the $V_{GS}$ sweep direction was not reported (Table 1).

In this work, we describe a pulsed measurement technique to suppress hysteresis in single-wall CNT FET transfer characteristics, and subsequently use it to extract effective mobility values without gate screening effects. The approach is quite general and could be applied to CNTs on other dielectrics, substrates, polymers, or to other nanoscale conductors (e.g. graphene) where unwanted hysteretic behavior is often observed. We find that increased off times between gate voltage pulses reduce measured hysteresis, and the transfer characteristics move towards a common, unique curve revealing a single value for the device mobility. By varying the pulse width and duty cycle in our measurements over a wide range (1 ms – 10 s), we also extract the relaxation times associated with environmental charge trapping at various temperatures from 80 – 453 K, in air and in vacuum. We adapt a tunneling front model [15-17] to extract the associated trap depths affecting hysteresis in our measurements. Finally, we investigate the error in ex-tracted carrier mobility in CNTs between the (unique) pulsed and (ambiguous) DC gate voltage measurements.

## 2. Experimental Methods

To fabricate the devices used in this study we begin by removing the native oxide from a bare highly doped (p+) Si wafer in a HF solution, followed by a 15 min clean in a 7:1 $H_2O_2$:$H_2SO_4$ (Piranha) solution. Approximately 70 nm of dry thermal $SiO_2$ is grown at 1150 °C. Next, ~2 Å Fe catalyst is deposited onto lithographically defined areas (~5×5 μm) by electron-beam (e-beam) evaporation. Carbon nanotubes are grown in an Atomate chemical vapor deposition (CVD) system by annealing the substrate at 900 °C in an Ar environment for 30 minutes, followed by CNT growth at 900 °C under $CH_4$, $C_2H_4$ and $H_2$ flow (~50:1:30). Metal pads are lithographically aligned to the pre-patterned catalyst and deposited by e-beam evaporation (1 nm Ti/ 40 nm Pd). The electrode pads are defined by lift-off in MicroChem Remover PG. The contacts are annealed at 300 °C in an Ar environment for 30 minutes. The highly doped (p+) silicon substrate served as the back gate [18], and CNTs were exposed to ambient from above, as shown in figure 1.

CNT diameter ($d$) and length ($L$) were measured by atomic force microscopy (AFM) and scanning electron microscopy (SEM), as shown in the supporting information figure S1 and the inset of figure 1(a). Transfer characteristics were measured using a Keithley 2612 dual-source measuring unit, at constant $V_{DS} = 50$ mV, while performing a pulsed sweep of $V_{GS}$ between ±10 V, see inset of figure 1 (b). Pulsed $I_D$-$V_{GS}$ characterization of CNT FETs is achieved through a custom script written in the Lua language, which is based on the Keithley 2612 instrument default $I_D$-$V_{GS}$ characterization script. The script has been made available for download on our web site [19]. The user-defined $V_{GS}$ sweep is applied in a pulsed linear fashion with a base voltage of $V_{GS} = 0$ V. Communication with the instrument is achieved through a LabView interface and the model KUSB-488A IEEE-488.2 USB-to-GPIB interface adapter. The gate voltage pulse period was varied from 2 ms – 10 s with the pulse width held constant at 1 ms. A constant pulse width was used because no significant dependence of hysteresis on it was found in the range of 250 μs – 1 ms. Measurements were made under varying conditions and temperatures. The devices in this study had diameters ranging from $d \approx 1.6 – 3.8$ nm and channel lengths $L \approx 2 – 7.5$ μm (see supporting information figure S2).

## 3. Results

The hysteresis gap ($\Delta V_T$) is defined as the difference in threshold voltage between the forward and backward $V_{GS}$ sweeps, as determined by the linear extrapolation method and illustrated in figure 2(a) [20]. Hysteresis dependence of pulsed measurements is compared in air and vacuum (~$10^{-5}$ Torr) at room temperature for two CNTs with similar length and diameters $d \approx 2.1$ nm (figures 2(a) and (b)) and $d \approx 1.7$ nm (figures 2(c) and (d)). Hysteresis is found to be reduced by increasing the length of the pulse off time ($t_{OFF}$). In air hysteresis is reduced by up to 75% (figure 2(a)) when $t_{OFF}$ is increased from 1 ms to 10 s. In vacuum hysteresis is nearly eliminated (figure 2(d)) when $t_{OFF}$ is increased from 1 ms to 10 s. Furthermore, hysteresis reduction in vacuum is more pronounced at shorter off times for the device with $d \approx 2.1$ nm, suggesting that charge injection into the substrate affects hysteresis less than charge trapping by surrounding water molecules (which partially desorb in vacuum) for this device [5]. However, for the device with $d \approx 1.7$ nm the exposure to vacuum has no effect on the hysteresis at shorter off times, possibly due to reduced surface area for water adsorption and the increased electric field (which scales roughly as ~$1/d$) at the CNT/SiO$_2$ interface. For this device, charge injection into the substrate is most likely the dominant cause of hysteresis.

Figure 3(a) shows measurements made in air at temperatures from 293 – 453 K, indicating the rate of hysteresis reduction ($\Delta V_T$) with $t_{OFF}$ increases with temperature. This suggests reduced charge trapping by the surrounding water molecules, and faster relaxation times of trapped charge at higher temperature. At low temperature in vacuum (80 K, in figure 3(a) inset) we find hysteresis is nearly constant at $\Delta V_T \approx 1.5$ V, similar to the behavior observed with DC measurements by Vijayaraghavan *et al*. [9] Figure 3(b) illustrates the dependence of $\Delta V_T$ on $t_{OFF}$ at room temperature in air and under vacuum. In both figures 3(a) (in air) and 3(b) (in vacuum) at short $t_{OFF}$ (< 100 ms), there is no significant dependence of $\Delta V_T$ on $t_{OFF}$. However, at higher $t_{OFF}$ there is a rapid decrease in hysteresis as the trapped charge surrounding the CNT has adequate time to relax during the off part of the gate voltage pulses. This indicates the typical relaxation (detrapping) times of injected charge into the substrate are greater than 100 ms.

## 4. Discussion

We can gain insight into the distribution of trap depths affecting hysteresis, i.e. those with tunneling times approximately between $0.01 - 10$ s, by numerically examining the charge tunneling and trapping process. We first estimate the electric field from the CNT into the $SiO_2$:

$$F\left(x\right) = \frac{V_{GS}}{x \ln\left(2\frac{t_{OX}}{r}\right)} \quad ; \qquad x \geq r \tag{1}$$

where $t_{OX}$ is the $SiO_2$ thickness, $r$ is the CNT radius, and $x$ is the distance from the center of the CNT into the $SiO_2$ [21]. Unlike in a parallel plate capacitor where the electric field is constant, this field can be very high near the $CNT/SiO_2$ interface given the extremely small CNT radius, even for only a few Volts applied across the $SiO_2$ dielectric. The band edge diagram of the $CNT/SiO_2$ interface is schematically displayed in the figure 4(a) inset. The barrier height associated with tunneling, $\Phi$, depends on CNT diameter through

$$\Phi \approx \varphi_{\mathrm{CNT}} - \chi_{\mathrm{SiO_2}} - E_G / 2 \tag{2}$$

where $\varphi_{\mathrm{CNT}} \approx 4.7$ eV is the CNT work function, $\chi_{\mathrm{SiO2}} \approx 0.95$ eV is the $SiO_2$ electron affinity [22], and $E_G \approx 0.84/d$ is the CNT band gap with the diameter $d$ given in nanometers [18]. The tunneling time constant can be written as

$$\tau = \tau_0 \exp\left\{ \int_{r}^{x_D+r} \frac{2m^* x'^{1/2}}{\hbar} \left[ \frac{\Phi}{x'} - qF\left(x'\right) \ln\left(\frac{x'}{r}\right) \right]^{1/2} dx' \right\} \tag{3}$$

where $m^* \approx 0.42m_0$ is the effective tunneling mass in $SiO_2$, $x_D$ is the trap depth, $m_0$ and $q$ are the electron mass and charge, respectively, and $\tau_0 \approx 6.6 \times 10^{-14}$ s is a characteristic time constant fitted against previous tunneling front model experiments in $SiO_2$ [15-17]. From equation 3 we can see that as $x_D$ approaches the $CNT/SiO_2$ interface, the time scale $\tau$ approaches $\tau_0$.

The effective potential ($V_{GS,eff}$) experienced by the CNT can in practice be different from that applied to the gate electrode. This is in part due to charge screening by adsorbed water molecules on the surface of the $CNT/SiO_2$, and to the injected charge during measurements. Therefore, the simple model described in equations 1-3 above is used to estimate the upper bounds of the trap depths ($x_D$) associated with relaxation times between $\tau = 0.01$ and 10 s [23, 24]. This is shown in figure 4 for CNTs of diameter $d = 1$ and 4 nm with an effective potential $V_{GS,eff} = 1$ and 5 V. As expected, the field is greater for the smaller diameter tube near the CNT/SiO$_2$ interface $(x - r = 0)$, shown in figure 4(a). As a result we expect CNTs of smaller diameter to populate traps further away from the CNT/SiO$_2$ interface, as shown in figure 4(b). Using this model we estimate the trap depths for the time constants $\tau = 0.01$ and 10 s to correspond roughly to $x_D \approx 4$ and 5 nm respectively, for a CNT FET with $d = 4$ nm at $V_{GS,eff} = 1$ V. For a CNT FET with $d = 1$ nm and $V_{GS,eff} = 5$ V the corresponding trap depths for time constants $\tau = 0.01$ and 10 s are $x_D \approx 6$ and 8 nm respectively. As the trap depth approaches the CNT/SiO$_2$ interface the model converges to $\tau_0$ for all cases. The model also suggests a dependence of measured hysteresis on CNT diameter. However, experimentally we do not find a clear dependence of hysteresis on either CNT diameter or length after comparing $\Delta V_T$ from the DC transfer characteristics of nineteen CNT FETs (see supporting information figure S2). We attribute this to variability in the SiO$_2$ surface roughness between different samples [25], to defects in the CNTs measured, and to ambient conditions which cannot be precisely controlled at the atomic scale of the CNT/SiO$_2$ interface during measurement. However, it is evident that the pulsed measurements described in this work yield consistent, reproducible results (i.e. hysteresis reduction) *in spite* of such variability between CNT samples, and the relatively straightforward approach should make it applicable to a wide range of nanostructures with inherent variability, such as graphene, nanowires, or molecular electronics.

We note the direction of the hysteresis collapse may provide some insight into the type of trap sites being populated. For example, hysteresis collapse towards more positive gate voltage and the reverse DC sweep (figures 2 (a) and (b)) could be indicative of hole traps depopulating. Hysteresis collapse towards the middle of the DC forward/backward sweeps (figures 2 (c) and (d)) could indicate an equal number of hole and electron traps depopulating. Hysteresis collapse toward negative gate voltages could indicate electron traps depopulating. In addition, we note that typical oxides have trap densities ranging from $10^{10} - 10^{13}$ cm$^{-2}$ [26] which correspond to only $1 - 600$ traps for typical CNTs in our study (~3 $\mu$m length and ~2 nm diameter). Thus, variation in the oxide quality on our test chips can strongly influence the electrical properties of CNT devices (also underscored by the lack of clear trends in figure S2).

## 5. Mobility Extraction

Before concluding, we compare the effective mobility extracted from the forward and reverse DC sweeps in air, with the mobility extracted from pulsed measurements with $t_{OFF}$ = 10 s under vacuum. This is done for the devices with similar length and diameters $d \approx 1.7$ nm and 2.1 nm in figure 5. The effective mobility is obtained as $\mu_{EFF} = GL/(qn)$ where $n = C'/[q(V_T-V_{GS})]$ is the carrier density per unit length obtained from the experimental data, $G = I_D/(V_{DS}-I_D R_C)$ is the drain conductance at $V_{DS}$ = 50 mV, and $C' = 2\pi\varepsilon/\ln(2t_{OX}/r)$ is the CNT capacitance per unit length with $\varepsilon \approx 2.2\varepsilon_0$ for CNTs on SiO$_2$ to effectively account for fringing fields [21]. $R_C$ is the contact resistance, estimated from measurements at low field ($R_{LF}$) such that $R_C = R_{LF}-R_0$, where $R_0$ is the intrinsic resistance of the CNT which depends on $L$ and the acoustic phonon mean free path, $\lambda_{AP} \approx 280d$ as described in our recent work [27]. For the device with $d \approx 1.7$ nm and $L \approx 2.6$ μm we obtain $R_0 \approx 42$ kΩ and for the device with $d \approx 2.1$ nm and $L \approx 2.5$ μm we obtain $R_0 \approx 34$ kΩ. The threshold voltage $V_T$ used in calculating $\mu_{EFF}$ is determined here by finding the gate voltage at a specified threshold drain current ($I_T$), such that $I_T \approx G/G_0 < 0.001$, where $G_0 = 4q^2/h$ is the quantum conductance of four CNT channels [27].

We find that at longer pulse $t_{OFF}$ times there is less discrepancy between forward and backward sweeps, and the extracted mobility approaches a common value, as shown in figures 5 (c) and (d). Moreover, we find the extracted mobility varies by approximately a factor of two between the forward and backward DC sweeps in air, highlighting the inadequacy of extracting mobility from a DC sweep. However, when measured with the pulsed technique in vacuum, the error in extracted mobility between the forward and backward $V_{GS}$ sweep is reduced to approximately 10% for the device with $d \approx 2.1$ nm and completely eliminated in the case of the device with $d \approx 1.7$ nm. It is interesting to note that the extracted $\mu_{EFF}$ from the pulsed measurement technique lies between that extracted from the forward and reverse DC sweeps. This suggests that Coulomb scattering due to trapped charge has a weaker effect on the CNT mobility than acoustic phonon scattering. Furthermore, we note that in both cases the mobility initially increases and then decreases with carrier concentration ($n$), peaking at $n \approx 0.5–1$ carriers/nm. This is precisely consistent with the inverse dependence of CNT mobility on the density of states (DOS), as the DOS first decreases when the Fermi level ($E_F$) moves away from the edge of the first sub-band, and then increases as $E_F$ enters the second sub-band, leading to a decrease in mobility as a new scattering channel becomes available. A quantitative model for the behavior of CNT effective mobility in the presence of multiple sub-band conduction was recently given by our work in Ref. [27]. This is particularly evident for the device with $d \approx 1.7$ nm and correlates well to the observed peak in the $I_D$-$V_{GS}$ characteristics in figures 5(b) and 5(d).

## 6. Conclusions

We have described a pulsed measurement method which eliminates unwanted hysteresis of CNT devices in air and under vacuum conditions. By varying the off time between gate voltage pulses we find the relaxation time of the trapped charge affecting hysteresis to be between 100 ms – 10 s. We also present a simple tunneling front model to extract the upper bounds of the charge trap depths, estimated to between 4 – 8 nm for CNTs of diameter 4 nm and 1 nm, respectively. The effect of hysteresis on mobility extractions from the forward and reverse DC gate voltage sweeps is determined, and it is shown that long pulse intervals at high temperature and under vacuum result in the extraction of a more consistent mobility value for CNTs. The approach presented here opens the door and could also be applied for more careful evaluations of other nanostructures with inherent variability and trapped charge effects, including graphene, nanowires, and molecular devices.

**Acknowledgements**

The authors thank the research staff at the Micro and Nanotechnology Laboratory (MNTL) for technical assistance, and Prof. J. Lyding (UIUC), Dr. J. Suehle, and Dr. C. Richter (NIST) for excellent suggestions. This work has been supported in part by the NASA KSC grant NNX08AL96G, the Nanoelectronics Research Initiative (NRI) MIND center, and the NSF CCF-0829907 grant. D.E. acknowledges support by the Micron Technology Foundation, the NSF and NDSEG Fellowships. S.D. acknowledges support by the NASA Aeronautics Scholarship. A.L. acknowledges support by the IBM and NRI Hans J. Coufal Fellowship.

**Table 1.** Mobility values reported for various CNTs in the literature.

| μ (cm$^2$V$^{-1}$s$^{-1}$) | d (nm) | L (μm) | V$_{GS}$ Sweep or Hysteresis Reduction Method |
|---|---|---|---|
| ~Ballistic | 3 | 0.3 | PMMA passivated [28] |
| 79,000 ± 8,000 | 3.9 | 325 | Reverse sweep [1] |
| 5,000 - 20,000 | <5 | 4000 | Not reported [29] |
| 16,000 | 4 | 4 | Forward [2] |
| 4,000 | 3 | 3 | PMMA passivated [28] |
| 2,500 | 1.5 | 10 | Forward sweep [2] |
| 1,000 - 4,000 | 1 to 4 | 1 to 3 | Vacuum [30] |
| 20 | 1.6 | 0.3 | Not reported [31] |
| 600 - 8,000 | Not reported | 3 | PEI doped [32] |

*Polymer coatings or vacuum conditions have sometimes been used to reduce hysteresis when extracting mobility [28, 30, 32]. In a few studies the direction of the sweep used for mobility calculation is unavailable [29, 31].

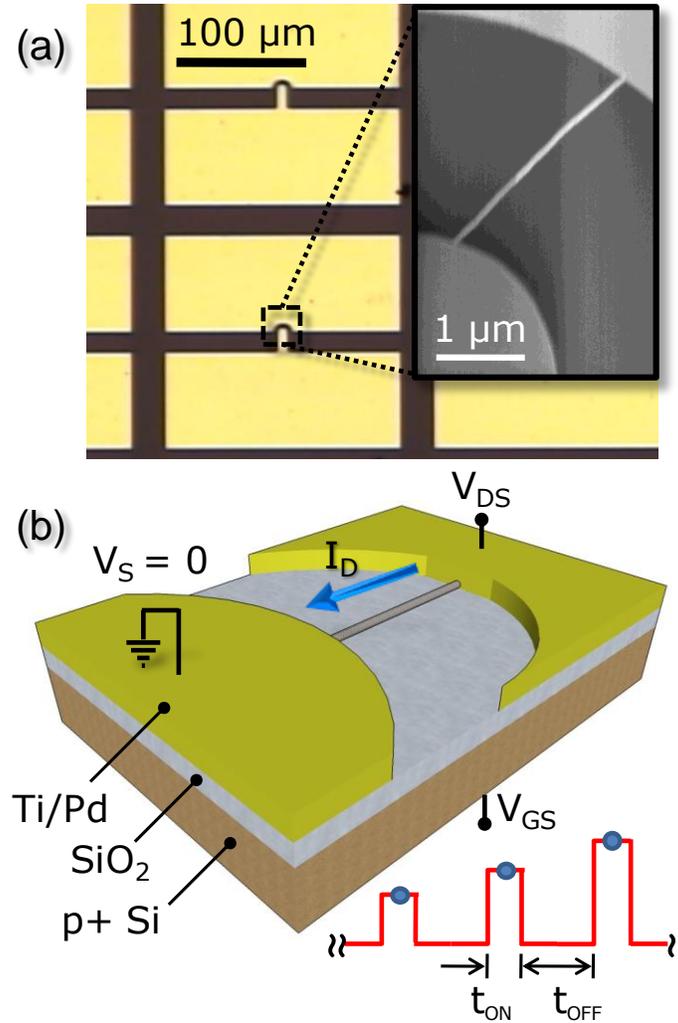

**Figure 1: (a)** Top view optical image of typical CNT devices used in this work. Semi-circular electrodes are adopted for tighter control of nanotube device length [18]. Inset shows SEM image of typical device. **(b)** Schematic of CNT test device and pulsed gate voltage train [19].

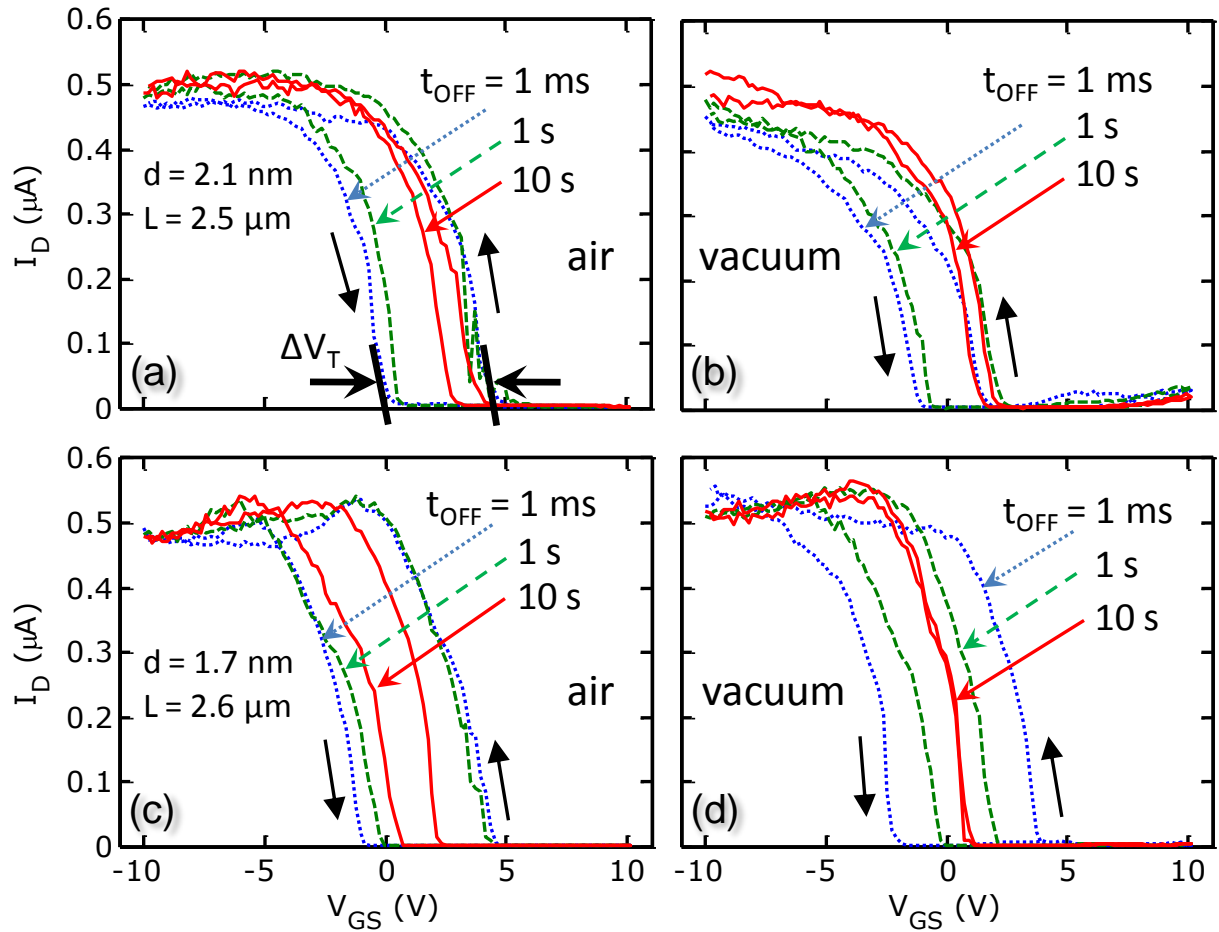

**Figure 2: (a)** Typical $I_D$-$V_{GS}$ transfer curves for a device with $d \approx 2.1$ nm in air and **(b)** in vacuum ($\sim 10^{-5}$ torr) at room temperature. The hysteresis gap ($\Delta V_T$) is defined as the difference between the forward and reverse sweep threshold voltage. The hysteresis loop indicates charge trapping into the substrate [11]. **(c)** Typical $I_D$-$V_{GS}$ transfer curves for a device with $d \approx 1.7$ nm in air and **(d)** in vacuum at room temperature. In all cases hysteresis is reduced by increasing $t_{OFF}$ of the applied $V_{GS}$ pulses.

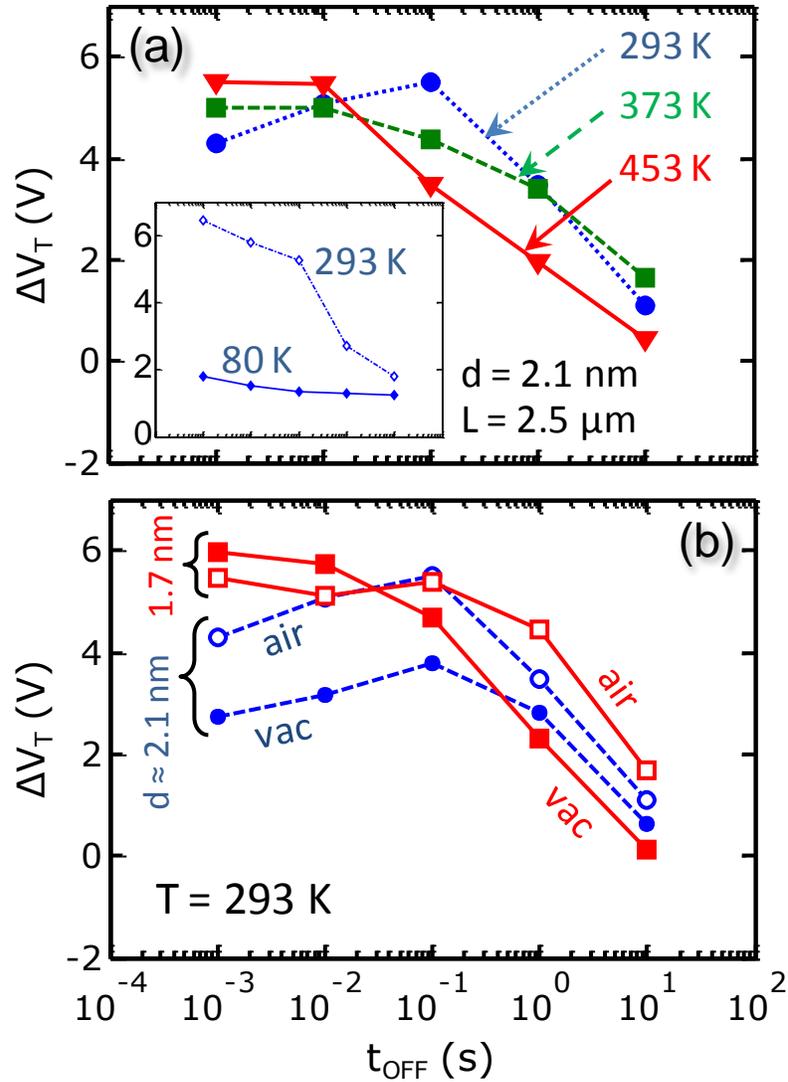

**Figure 3: (a)** Hysteresis gap ($\Delta V_T$) vs. pulse off-time ($t_{OFF}$) for the device in figure 2(a) at temperatures of 293 K (●), 373 K (■), and 453 K (▼) in air. Pulsed measurements are more effective in reducing the hysteresis at higher temperatures. Inset shows nearly constant $\Delta V_T \approx 1.5$ V with various $t_{OFF}$ in vacuum at low temperature (80 K). Also see figure S3 in the supporting information. **(b)** $\Delta V_T$ vs. $t_{OFF}$ for the devices in figures 2(a) and 2(b). For both the hysteresis reduction is greatest at $t_{OFF} > 100$ ms, indicative of relatively long trap relaxation times.

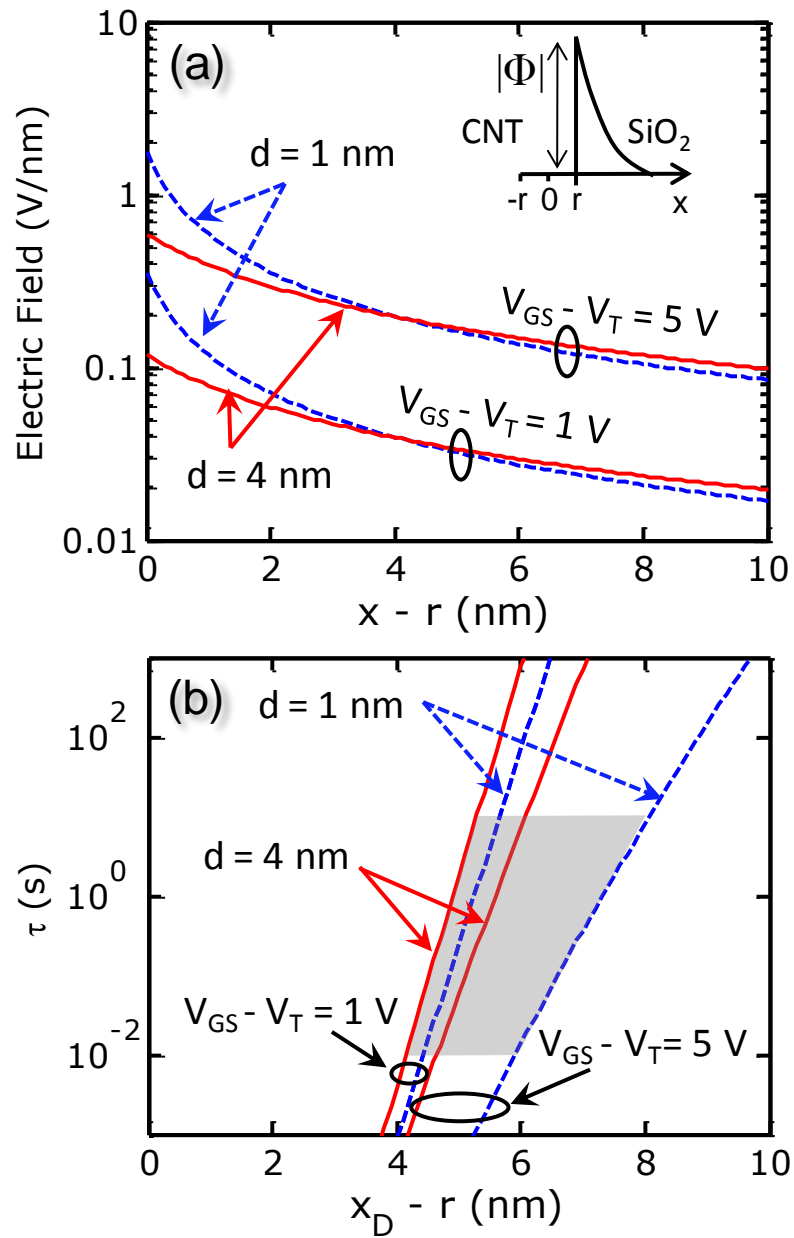

**Figure 4: (a)** Calculated electric field near the CNT/SiO₂ interface for CNTs of diameter $d \approx 1$ nm (dashed blue) and $\approx 4$ nm (solid red line) at gate voltage overdrive $V_{GS}$-$V_T \approx 1$ and 5 V. **(b)** Calculated tunneling time vs. trap depth from the CNT/SiO₂ interface for CNTs of $d \approx 1$ and 4 nm at $V_{GS}$-$V_T \approx 1$ and 5 V. The estimated trap depth window affecting hysteresis in our measurements is shown as the shadowed region, corresponding to $0.01 - 10$ s time scales.

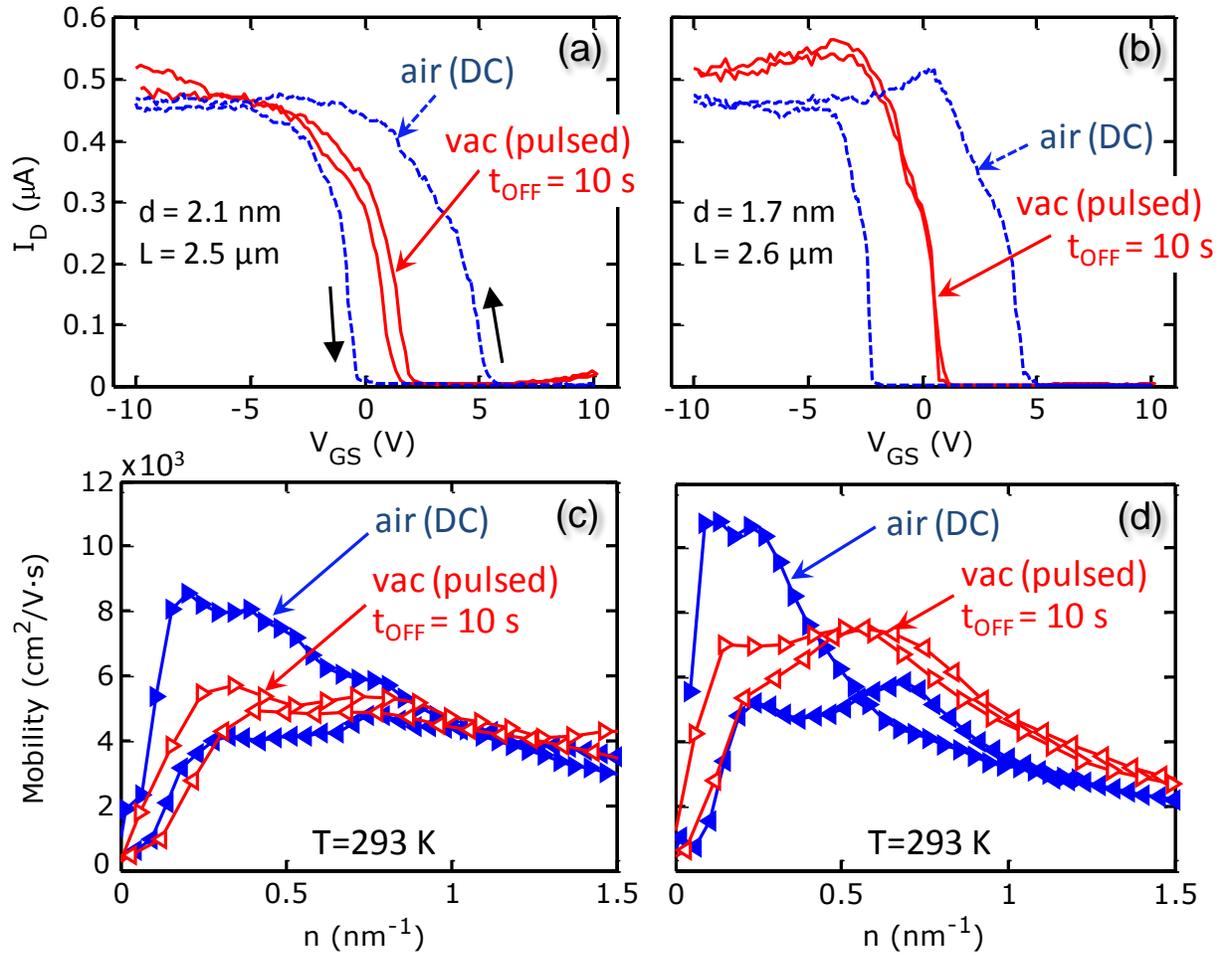

**Figure 5: (a)** Comparison of DC transfer curves in air (dashed) and pulsed under vacuum conditions (solid) for the device with diameter $d \approx 2.1$ nm. **(b)** Similar data for a device with diameter $d \approx 1.7$ nm in air (dashed) and pulsed under vacuum conditions (solid). **(c)** Corresponding mobility extraction for the device in (a) and **(d)** for the device in (b). Rightward triangles indicate mobility from forward $V_{GS}$ sweep and leftward triangles from reverse sweep. Filled triangles indicate mobility from DC $V_{GS}$ sweeps and open triangles from pulsed $V_{GS}$ sweeps.

**Supporting Information**

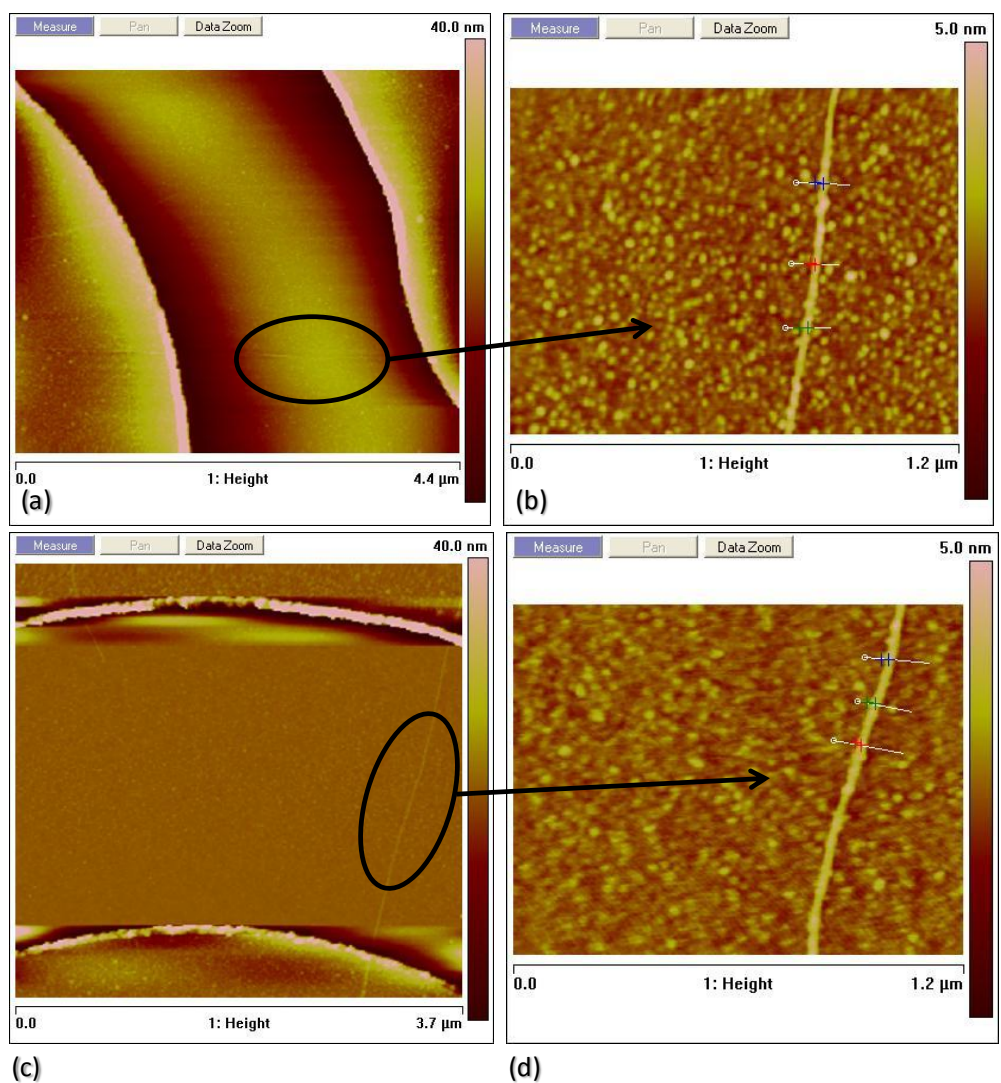

**Figure S1: (a, b)** AFM images of a CNT FET with one active device connection corresponding to the device characteristics shown in figures 2 (a) and (b) and 5 (a) and (c). The CNT in the channel region of this device has $d \approx 2.1$ nm and $L \approx 2.5$ μm. Reported diameters are an average of three separate height measurements, as shown in figure S1 (b). **(c, d)** AFM images of a CNT FET with one active device connection corresponding the device characteristics shown in figures 2 (c) and (d) and 5 (b) and (d). The CNT in the channel region of this device has $d \approx 1.7$ nm and $L \approx 2.6$ μm. Reported diameters are an average of three separate height measurements, as shown in figure S1 (d).

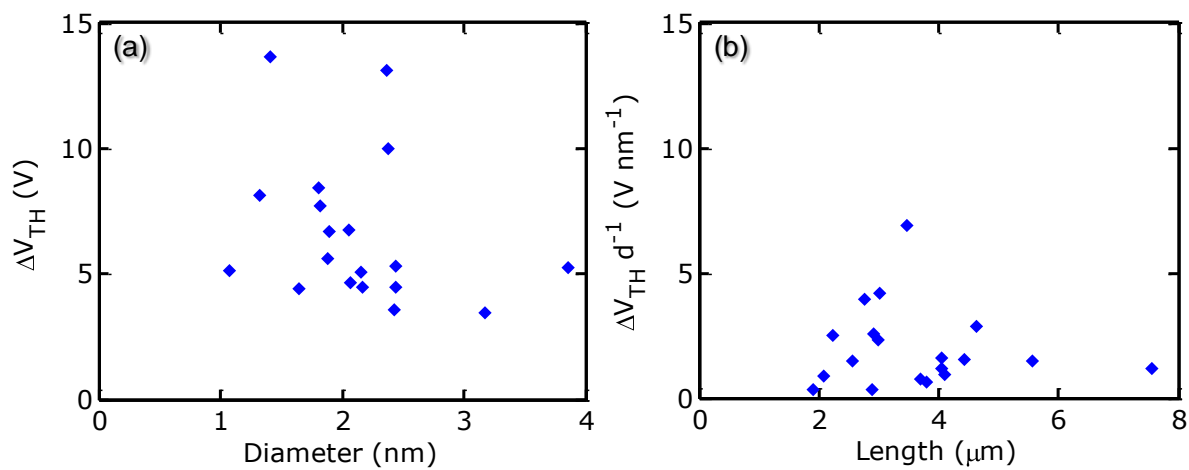

**Figure S2:** **(a)** Hysteresis gap ($\Delta V_T$) vs. diameter for multiple CNT FETs. $\Delta V_T$ is extracted from the DC transfer characteristic measured in air at room temperature for all devices. No apparent dependence of hysteresis on CNT diameter is found. **(b)** $\Delta V_T$ (normalized by CNT diameter $d$) vs. length $L$ for multiple CNT FETs. No apparent dependence on CNT length is found.

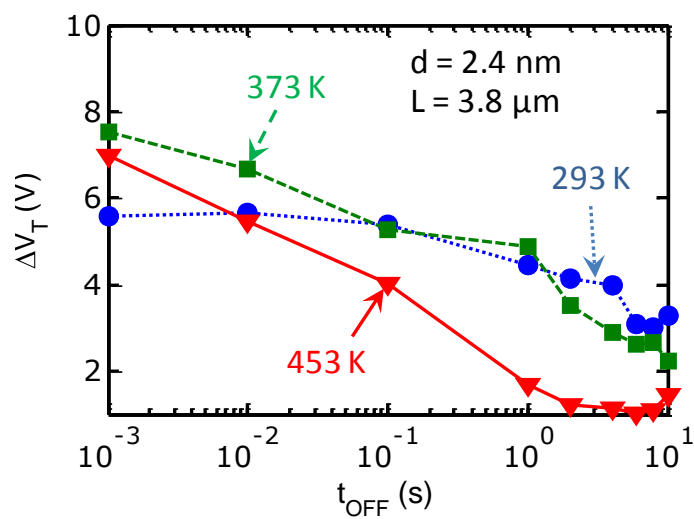

**Figure S3:** Hysteresis gap ($\Delta V_T$) vs. pulse off-time ($t_{OFF}$) for an additional device at 293 K (●), 373 K (■), and 453 K (▼) in air. Pulsed measurements are once again found to be more effective in reducing the hysteresis at higher temperatures, as in figure 3 (a). We have tested ten CNT devices in total with the pulsed measurement method, and found similar trends for all, except for one that was covered in significant photoresist (PR) residue by AFM inspection.